%% file: mypaper.tex
\pdfoutput=1

\documentclass[11pt]{article}

\usepackage{emnlp2021}

\usepackage{xr}
\makeatletter
\newcommand*{\addFileDependency}[1]{
  \typeout{(#1)}
  \@addtofilelist{#1}
  \IfFileExists{#1}{}{\typeout{No file #1.}}
}
\makeatother

\newcommand*{\myexternaldocument}[1]{
    \externaldocument{#1}
    \addFileDependency{#1.tex}
    \addFileDependency{#1.aux}
}

\myexternaldocument{appendix}

\usepackage{times}
\usepackage{latexsym}
\usepackage{nert}
\usepackage{comment}
\usepackage{footmisc}
\usepackage{soul}
\usepackage{multicol}
\usepackage{color, colortbl}
\definecolor{myblue}{RGB}{0, 255, 255}
\definecolor{myred}{RGB}{255, 0, 0}
\definecolor{myyellow}{RGB}{255, 255, 0}

\DeclareRobustCommand{\hlblue}[1]{{\sethlcolor{myblue}\hl{#1}}}
\DeclareRobustCommand{\hlyellow}[1]{{\sethlcolor{myyellow}\hl{#1}}}

\DeclareRobustCommand{\hlorange}[1]{{\sethlcolor{orange}\hl{#1}}}

\usepackage{refcount}

\newcommand{\myfootnotetext}[1]{\footnotetext{#1\label{fn:text}%
        \edef\fnmark{\getpagerefnumber{fn:mark}}%
        \edef\fntext{\getpagerefnumber{fn:text}}%
        \ifx\fnmark\fntext\else\ClassWarning{}{footnote mark and text on different pages!}\fi}}

\usepackage{footnote}
\makesavenoteenv{tabular}
\makesavenoteenv{table}

\usepackage[T1]{fontenc}

\usepackage{microtype}

\title{How Does Counterfactually Augmented Data Impact Models for Social Computing Constructs?}

\author{Indira Sen \quad Mattia Samory \quad Fabian Flöck\\ 
  GESIS – Leibniz Institute for the Social Sciences \\
  \{\emldisplay{indira.sen@gesis.org}{indira.sen},
  \emldisplay{mattia.samory@gesis.org}{mattia.samory}, \emldisplay{fabian.floeck@gesis.org}{fabian.floeck}\}\texttt{@gesis.org} \\
  \AND
Claudia Wagner \\
  GESIS and  RWTH Aachen University \\
   \eml{claudia.wagner@gesis.org} 
  \And
Isabelle Augenstein \\
  University of Copenhagen\\
   \eml{augenstein@di.ku.dk} \\
   \\
}
\date{}

\begin{document}
\maketitle
\begin{abstract}

As NLP models are increasingly deployed in socially situated settings such as online abusive content detection, it is crucial to ensure that these models are robust. One way of improving model robustness is to generate counterfactually augmented data (CAD) for training models that can better learn to distinguish between core features and data artifacts. While models trained on this type of data have shown promising out-of-domain generalizability, it is still unclear what the sources of such improvements are. We investigate the benefits of CAD for social NLP models by focusing on three social computing constructs --- sentiment, sexism, and hate speech. Assessing the performance of models trained with and without CAD across different types of datasets, we find that while models trained on CAD show lower in-domain performance, they generalize better out-of-domain. We unpack this apparent discrepancy using machine explanations and find that CAD reduces model reliance on spurious features. 
Leveraging a novel typology of CAD to analyze their relationship with model performance, we find that CAD which acts on the construct directly or a diverse set of CAD leads to higher performance. 

\end{abstract}

\section{Introduction}

\begin{table}
\small
\centering
\begin{tabular}{@{}lll@{}}
\toprule
construct                                              & original                                                                                               & counterfactual                                                                                           \\ \midrule
\begin{tabular}[c]{@{}l@{}}sentiment\end{tabular}  & \begin{tabular}[c]{@{}l@{}}I thought this movie \\ was very \hlblue{well} \\ put together.\end{tabular}         & \begin{tabular}[c]{@{}l@{}}I thought this movie \\ was very \hlblue{haphazardly} \\ put together.\end{tabular} \\
\midrule
sexism                                                 & \begin{tabular}[c]{@{}l@{}}\hlorange{Females} should not \\ commentate on sport\end{tabular}                   & \begin{tabular}[c]{@{}l@{}}\hlorange{AI} should not \\ commentate on sport\end{tabular}                          \\
\midrule
\begin{tabular}[c]{@{}l@{}}hate speech\end{tabular} & \begin{tabular}[c]{@{}l@{}}Lets talk about the \\ antithesis of hard \\ work: \hlyellow{immigrants}\end{tabular} & \begin{tabular}[c]{@{}l@{}}Lets talk about the \\ antithesis of hard \\ work: \hlyellow{my brother}\end{tabular}   \\ \bottomrule
\end{tabular}
\caption{\textbf{Examples of original/counterfactual pairs for sentiment, sexism, and hate speech.} As pairs of data with minimal textual differences (color-coded here) but different labels, counterfactual examples can improve NLP models' focus on consequential features of the construct over dataset artifacts.}
\label{tab:counterfactual_intro}
\end{table}

Dataset design is receiving increasing attention, especially in response to concerns related to the generalizability of machine learning-based NLP models. Recent critiques argue that models trained for NLP tasks may end up ``learning the dataset'' rather than a particular \textit{construct}~\cite{bras2020adversarial}, i.e, the intangible measure like sentiment or stance 
that is the ultimate goal of the learning task~\cite{jacobs2021measurement}. In particular, in the process of inferring the mapping between an input space and output space, models may learn cues in the dataset which are spuriously correlated with the construct~\cite{schlangen2020targeting}. For example, sentiment models trained on movie reviews tend to learn more about movies than about sentiment, thereby failing to measure it as accurately in e.g., news media~\cite{puschmann2018turning}. This potential learning of spurious cues over meaningful manifestations of the construct makes it especially difficult to foresee how even small differences in the context of deployment would affect the performance of NLP models, with undesirable consequences for their applicability at large. The issue of model robustness is all the more crucial for social computing NLP models, particularly for constructs like hate speech and sexism, which are often deployed in detecting abusive content on online platforms~\cite{perspective2021}. In such settings, there is a risk of high societal and human harms such as sanctioning marginalized voices due to model misclassification and bias~\cite{guynn2019facebook}. Even in contexts other than online governance, such as using social NLP models for detecting abuse faced by a certain subpopulations on a particular online space, we incur the risks and consequences of mismeasurement~\cite{pine2015politics,wagner2021measuring}.

One suggested solution to address the issue of spurious features is counterfactually augmented data (CAD)---instances generated by human annotators that are minimally edited to flip their label---and their variations such as iterative benchmark design~\cite{potts2020dynasent}, contrast data generation~\cite{gardner2020evaluating},\footnote{Counterfactually augmented data and contrast sets refer to the same concept: making minimal changes to flip labels but have different conceptual grounding---causality for CAD and modeling decision boundaries for constrast sets.} and their combination~\cite{vidgen2020learning}. Drawing on the rich history of counterfactuals~\cite{pearl2018causal,lewis2013counterfactuals,kasirzadeh2021use}, the promise of CAD is to offer a causality-based framework where only cues that are meaningfully associated with the construct are edited --- which is expected to be conducive to models learning less spurious features. Indeed, recent work has shown that models trained on CAD generalize better out of domain
~\cite{kaushik2019learning,samory2021call}. Yet, it is not well understood why or how these counterfactuals are effective, especially for social NLP tasks--- \emph{do they reduce dependence on spurious features and to what extent?}
 
\textbf{This work. } We analyze how CAD affects social NLP models. Unlike previous work, we leverage multiple, related social computing constructs to avoid confounds that may arise due to the specific settings of a single construct. We conduct our experiments on three text classification tasks: sentiment, sexism, and hate speech identification. Sentiment has been thoroughly analyzed in past NLP robustness work, and abusive content has been widely studied in NLP~\cite{schmidt2017survey,vidgen2020directions,jurgens2019just,sarwar2021flagging}. However, sexism and hate speech have not been studied in as much detail in the specific context of the impact of training on CAD. The multifaceted nature of these constructs warrants further investigation, especially in the context of developing models with less spurious features. 

First, we ask: \textbf{(RQ1) do models trained on CAD outperform models trained on original, unaltered data? }We assess the overall performance of these two types of models and find that while models trained on original data outperform those trained on CAD in-domain, the opposite is true out-of-domain--- models trained on CAD are more robust out-of-domain. 

Next, we analyze \textbf{(RQ2) the characteristics of effective counterfactuals}, categorizing CAD according to their generation strategy, e.g., whether a negation was added or a gender word removed. Using this typology, we distinguish between \textbf{construct-driven} CAD, generated by directly acting on the construct (e.g., removing gender identity terms in sexism) versus \textbf{construct-agnostic} ones, generated by other strategies (e.g., negating a clause).
We find that construct-driven counterfactuals are more effective than construct-agnostic ones, especially for sexism.

We unpack the gain in out-of-domain performance by \textbf{analyzing (RQ3) whether models trained on CAD rely on less spurious features}. Complementing prior work, which has focused on the overall performance of models trained on CAD, we use explainability techniques to understand what models have learned. 
We find that models trained on CAD promote core, or non-spurious features, more than models not trained on CAD.

\paragraph{Overall contributions} Whereas previous work mainly assessed \textit{how much} CAD affects model performance, we focus on \textit{why} counterfactually augmented data improves performance for social computing NLP models. Our work has several implications on designing datasets and data augmentation, especially with respect to the benefits of different types of CAD. We release our code and collated data with the type of CAD labels for all three constructs to facilitate future research here: \url{https://github.com/gesiscss/socialCAD}.

\begin{table*}[]
    \centering
    \setlength{\tabcolsep}{6pt}
    \small
\begin{tabular}{@{}ccccccccccc@{}}
\toprule
\multirow{2}{*}{construct}   & \multicolumn{7}{c}{in-domain}                                                                                                                                                                                        & \multicolumn{3}{c}{out-of-domain}                                                        \\ \cmidrule(l){2-11} 
                             & reference         & \multicolumn{2}{c}{train}                                      & \multicolumn{2}{c}{counterfactual}                             & \multicolumn{2}{c}{test}                                       & reference               &        &                                                       \\ \midrule
\multirow{2}{*}{sentiment}   & \multirow{2}{*}{\citeauthor{kaushik2019learning}} & \textbf{pos}    & \textbf{neg}                     & \textbf{pos}    & \textbf{neg}                                                   & \textbf{pos}    & \textbf{neg}                                                   & \multirow{2}{*}{Kaggle\footnotemark[4]} & \textbf{pos}    & \textbf{neg}                                                   \\
                             &                   & 856    & 851                                                   & 851    & 856                                                   & 245    & 243                                                   &                         & 1103   & 1001                                                  \\
\midrule
\multirow{2}{*}{sexism}      & \multirow{2}{*}{\citeauthor{samory2021call}} & \textbf{sexist} & \begin{tabular}[c]{@{}c@{}}\textbf{not}\\ \textbf{-sexist}\end{tabular} & \textbf{sexist} & \begin{tabular}[c]{@{}c@{}}\textbf{not}\\ \textbf{-sexist}\end{tabular} & \textbf{sexist} & \begin{tabular}[c]{@{}c@{}}\textbf{not}\\ \textbf{-sexist}\end{tabular} & \multirow{2}{*}{EXIST\footnotemark[5]}  & \textbf{sexist} & \begin{tabular}[c]{@{}c@{}}\textbf{not}\\ \textbf{-sexist}\end{tabular} \\
                             &                   & 1244   & 1610                                                  & -      & 912                                                   & 534    & 690                                                   &                         & 1636   & 1800                                                  \\
\midrule
\multirow{2}{*}{hate speech} & \multirow{2}{*}{\citeauthor{vidgen2020learning}} & \textbf{hate}   & \begin{tabular}[c]{@{}c@{}}\textbf{not}\\ \textbf{-hate}\end{tabular}   & \textbf{hate}   & \begin{tabular}[c]{@{}c@{}}\textbf{not}\\ \textbf{-hate}\end{tabular}   & \textbf{hate}   & \begin{tabular}[c]{@{}c@{}}\textbf{not}\\ \textbf{-hate}\end{tabular}   & \multirow{2}{*}{\citeauthor{basile2019semeval}}       & \textbf{hate}   & \begin{tabular}[c]{@{}c@{}}\textbf{not}\\ \textbf{-hate}\end{tabular}   \\
                             &                   & 6524   & 5767                                                  & 5096   & 5852                                                  & 471    & 464                                                   &                         & 1260   & 1740                                                  \\ \bottomrule
\end{tabular}
\caption{\textbf{Constructs and datasets used in this work.} In-domain datasets are used for both training and testing, while out-of-domain datasets are exclusively used for testing. All in-domain datasets contain human-generated counterfactuals for both labels, except sexism where there is only counterfactual data for the negative class.} 
    \label{tab:data}
\end{table*}

\section{Training with Counterfactual Data}

\subsection{Motivation}

For a given text with an associated label, say a positive tweet, a counterfactual example is obtained by \textit{making minimal changes to the text in order to flip its label}, i.e., into a negative tweet.
Table~\ref{tab:counterfactual_intro} shows original-counterfactual pairs for the three types of NLP constructs studied in this paper.
Counterfactual examples in text have the interesting property that, since they were generated with minimal changes, they allow one to focus on the manifestation of the construct; in our example, what makes a tweet have positive sentiment. 

\subsection{Task Setting}

Formally, we have a model $f(x) = y$; $y$ is an application task label; $x$ is an instance that can be drawn from the original data set, or from the set of counterfactual data ($f(x_{c}) = \overline{y}$); $f$ is a learned feature representation. We optimise the binary cross entropy loss for $l(f, x, y)$ during learning.

There are different ways of incorporating counterfactuals; here, we simply treat them as ordinary training instances. This means any text classification model can be used for training on CAD. We learn feature representations on fully original data (non-counterfactual or \textbf{nCF} models) or on a combination of counterfactuals and original data (counterfactual or \textbf{CF} models). 

We have different sampling strategies --- random and stratified sampling in different proportions to ensure various counterfactual generation strategies are presented equally. To ensure fair comparison between CF and nCF models, we train both types of models on equal sized datasets --- for CF models, we simply \textit{substitute} a portion of the original data with CAD. We either randomly sample the CAD (RQ1, RQ3), or sample based on CAD type (RQ2).

\section{Experimental Setup}

\subsection{Datasets}\label{sec:data}

Table~\ref{tab:data} summarizes the datasets used in this work. 
\paragraph{In vs. out-of domain} We consider two types of non-synthetic datasets per construct --- in-domain (\textbf{ID}) and out-of-domain (\textbf{OOD}). Models are both trained and tested on in-domain data while out-of-domain data is fully held-out for testing. 
For the in-domain data, we use the same train-test splits as the original work, except for sexism, where a test set is not provided, so we do a stratified split of 70-30 (train-test). The out-of-domain data is exclusively used for testing. The EXIST data\footnotemark[5] also contains Spanish data, but we restrict ourselves to only English content in this work, as the in-domain data used for training is in English.

\paragraph{Counterfactually augmented data} All in-domain datasets we consider come with counterfactually augmented data, annotated by 
trained crowdworkers~\cite{kaushik2019learning,samory2021call} or expert annotators~\cite{vidgen2020learning}.\footnote{Only \citet{samory2021call} generate more than one counterfactual example per original, but to keep things consistent across all constructs, we randomly sample one counterfactual-original pair for sexism. \citet{vidgen2020learning} generate different types of synthetic data, including CAD, as a part of dynamic benchmarking for collecting hate speech data. We only use the original-counterfactual pairs from their dataset.}
Note that since previous work has shown that models trained on CAD tend to perform well on counterfactual examples~\cite{kaushik2019learning,samory2021call}, to prevent reporting inflated performance, we do not include counterfactual examples in any of the test sets.

Following~\citet{kaushik2019learning}, for sentiment and hate speech, the CF models are trained on 50\% original and 50\% CAD data, while for sexism, which has CAD only for non-sexist examples, models are trained on 50\% original sexist data, 25\% original non-sexist data, and 25\% counterfactual non-sexist data~\cite{samory2021call}.\footnote{We assess the effect of CAD proportion on model performance in Appendix \ref{app:injection}}

\footnotetext[4]{https://www.kaggle.com/c/tweet-sentiment-extraction This also contains tweets with neutral labels, but in this work, we restrict ourselves to positive and negative tweets only.}
\footnotetext[5]{From the EXIST 2021 shared task on sexism detection~\cite{EXIST2021} available at http://nlp.uned.es/exist2021/}

\paragraph{Adversarial test set} To further assess model robustness, in addition to evaluating on in-domain and out-of-domain data, we generate automated adversarial examples which do not flip the label through textattack~\cite{morris2020textattack}. These are of two types --- one which replaces words with synonyms (adv\_swap, \citet{wei2019eda}) and another which replaces named entities with other named entities (adv\_inv, \citet{ribeiro2020beyond}). They are both generated by perturbing the in-domain dataset. Note that due to the nature of these perturbations, adversarial data can only be generated for a subset of the training data, e.g., if an example does not contain any named entities, then we cannot generate an adv\_swap version of it.

\subsection{Text Classification Methods.}\label{sec:methods} 

We use two different text classification models: logistic regression (LR) and finetuned-BERT~\cite{devlin2018bert}. We do so as we want to contrast a basic model trained from scratch, which only learns simple features directly observed in the dataset (LR); and one which encodes a combination of background knowledge and application dataset knowledge, and is capable of learning complex inter-dependencies between features (BERT). We train LR with TF-IDF bag-of-words feature representations using sklearn~\cite{pedregosa2011scikit}, while the BERT base model is used for finetuning in conjunction with the subword tokenizer using HuggingFace Transformers~\cite{wolf2019huggingface}.

Each model is trained using 5-fold cross-validation and we use gridsearch for hyperparameter tuning. We conduct 5 runs for all models to reduce variance. We report the hyperparameters of all our models and their bounds in Appendix~\ref{app:hyper}.

\section{Experiments}

We first start by assessing overall performance on different types of data (RQ1), followed by introducing a typology of different types of CAD in order to understand if certain strategies of generating CAD are better for model performance (RQ2), and end by using explanations to understand which features the CF models promote (RQ3). 
Unless specified otherwise, we report results for BERT, while including the results for LR in the appendix for completeness (Appendix~\ref{app:lr_results}). We measure performance using macro F1 and positive class F1, where the latter metric is significant for constructs like sexism and hate speech.

\subsection{RQ1: Does CAD improve model performance?}\label{sec:rq1}

We compare the performance of the two types of models: trained on counterfactual data (CF) and trained on original data (nCF) on three different test sets: held-out in-domain test set, out-of-domain data, and adversarial examples. Table \ref{tab:id_od_performance} shows the results for in-domain and out-of-domain performance with CF and nCF data for BERT vs LR. Table \ref{tab:performance_adv} shows results with BERT for adversarial data. Recall that since we can only generate adversarial examples for a subset of the original data, we also include results on the original data for fair comparison. Results for LR models follow a similar trend and are included in Appendix \ref{app:lr_results}.

\begin{table}
\centering
\small
\setlength{\tabcolsep}{5pt}
\begin{tabular}{@{}lllrrrr@{}}
\toprule
                                                                       &        & mode    & \multicolumn{2}{c}{pos F1} & \multicolumn{2}{c}{macro F1} \\ \midrule
construct                                                              & method & dataset & CF           & nCF         & CF            & nCF          \\ \midrule
\multirow{4}{*}{\begin{tabular}[c]{@{}l@{}}sentiment\end{tabular}} & BERT   & ID*     & 0.85         & 0.89        & 0.85          & 0.89         \\
                                                                       &        & OOD*    & 0.87         & 0.85        & 0.85          & 0.83         \\
                                                                       & LR     & ID      & 0.82         & 0.86        & 0.81          & 0.86         \\
                                                                       &        & OOD*    & 0.77         & 0.71        & 0.74          & 0.58         \\ \midrule
\multirow{4}{*}{sexism}                                                & BERT   & ID*     & 0.80         & 0.82        & 0.81          & 0.84         \\
                                                                       &        & OOD*    & 0.62         & 0.42        & 0.66          & 0.56         \\
                                                                       & LR     & ID      & 0.69         & 0.75        & 0.72          & 0.79         \\
                                                                       &        & OOD*    & 0.43         & 0.32        & 0.55          & 0.50         \\ \midrule
\multirow{4}{*}{\begin{tabular}[c]{@{}l@{}}hate\\ speech\end{tabular}} & BERT   & ID*     & 0.93         & 0.98        & 0.93          & 0.98         \\
                                                                       &        & OOD*    & 0.62         & 0.58        & 0.66          & 0.63         \\
                                                                       & LR     & ID*     & 0.72         & 0.92        & 0.72          & 0.92         \\
                                                                       &        & OOD*    & 0.45         & 0.41        & 0.57          & 0.49         \\                                                                       
\bottomrule
\end{tabular}
\caption{\textbf{Model Performance (positive class precision and macro F1) averaged over 5 runs.} * indicates significant results (p < 0.01) in McNemar's Test. CF models outperform nCF models in out-of-domain data, while the opposite is true for in-domain data.}
\label{tab:id_od_performance}
\end{table}

\subsection{Results}

The overall results indicate that \textit{counterfactual models outperform non-counterfactual models on out-of-domain data, while results are  mixed for in-domain data.}. 
There are several possible explanations of this -- on one hand, the lower performance on the in-domain data could be due to the prevalence of spurious or domain-specific features in the nCF models as opposed to the CF models. On the other hand, CF models tend to learn less domain-specific features and more ‘general’ features, which leads to performance gains in other domains that the construct manifests in (as we explore in RQ3). 

As for adversarial data, it appears that CF models perform worse on it than their nCF counterparts in absolute terms. Note though that the adversarial data is automatically generated from the in-domain data, which indicates that nCF models have an advantage on it since nCF models might be picking up artifacts in the in-domain data that are also present in the adversarial examples (See Section~\ref{sec:data}). On the other hand, \textit{we do not find CF models' performance degrading on adversarial data anymore than nCF models, and in certain cases have smaller gaps between original and adversarial performance compared to their nCF counterparts (the case of adv\_inv)}, implying that CF models are equally robust, if not more.

\begin{table}
\small
\centering
\begin{tabular}{@{}lllllll@{}}
\toprule
          & \multicolumn{2}{l}{sentiment} & \multicolumn{2}{l}{sexism} & \multicolumn{2}{l}{hate speech} \\ \midrule
          & CF            & nCF           & CF           & nCF         & CF             & nCF            \\
\midrule
original  & 0.85          & 0.89          & 0.81         & 0.85        & 0.92           & 0.98           \\
adv\_inv  & 0.84          & 0.88          & 0.81         & 0.83        & 0.92           & 0.98           \\
\midrule
original  & 0.85          & 0.89          & 0.8          & 0.84        & 0.93           & 0.98           \\
adv\_swap & 0.81          & 0.83          & 0.76         & 0.78        & 0.85           & 0.96           \\ \bottomrule
\end{tabular}
\caption{\textbf{BERT Performance (macro F1) for adversarial data}; performance on the original in-domain subset added for comparison. As the adversarial data was generated from the original data, it is expected that nCF models have an advantage there, yet CF model performance does not degrade on the adversarial examples any more than on their nCF counterparts.}
\label{tab:performance_adv}
\end{table}

To summarize, we determine that CAD improves model robustness, especially for out-of-domain generalization. It neither helps nor hinders performance on adversarial examples. While the BERT models have much higher performance than LR, both family of models show similar trends. 

\subsection{RQ2: What are the characteristics of effective counterfactuals?}

\begin{table}[!t]
\small
\centering
\begin{tabular}{@{}lrrrll@{}}
\toprule
construct   & affect & gender & identity & neg  & hedge \\ \midrule
sentiment   & \textbf{0.98}   & 0.11   & 0.03     & 0.75 & 0.39   \\
sexism      & 0.18   & \textbf{0.79}   & 0.15     & 0.10 & 0.01   \\
hate speech & 0.55   & 0.21   & \textbf{0.23}     & 0.16 & 0.13   \\ \bottomrule
\end{tabular}
\caption{\textbf{The distribution of different modification strategies.} The proportions in bold refer to the construct-driven types --- affect for sentiment, gender for sexism, identity for hate speech.}
\label{tab:cf_type_dist}
\end{table}

Whereas the previous analyses assess whether CF models are more robust or not, we now turn to the question of whether all CAD is equally effective in improving classifier performance. Armed with a minimal set of instructions, annotators use several different strategies for generating CAD. Are some better than others? We aim to answer this question by categorizing different types of counterfactuals based on the strategy used to generate them. Then, to understand the `power' of different types of CAD, we assess the overall performance of models trained on the different types. 

\textbf{A Typology of Counterfactuals.} Previous work has manually assessed a sample of counterfactuals to understand the strategies used to generate them, such as introducing negation or distancing the speaker~\cite{kaushik2019learning,vidgen2020learning}. Yet, to the best of our knowledge, there is no categorization of the entire dataset of counterfactuals. Inspired by causal inference, particularly the notion of direct and indirect mediation~\cite{pearl2014interpretation,frolich2014direct}, we describe two distinct types of counterfactual data generation: \textit{construct-driven} and \textit{construct-agnostic}. Construct-driven CAD are generated by directly acting on the construct, e.g. replacing the gender word in sexism, or altering the affect-laden word in sentiment. On the other hand, construct-agnostic CAD are generated by indirectly acting on the construct, through general-purpose strategies such as introducing sarcasm or negation which yields CAD for several constructs (see Table~\ref{tab:cf_type_dist}). Since construct-driven CAD directly act on the construct, we hypothesize that \textit{construct-driven strategies are more effective.} 

To determine which instances represent which modification strategy, we use a simple lexicon-based automatic annotation strategy. Based on strategies manually assessed in previous literature~\cite{kaushik2019learning,vidgen2019challenges}, we devise 5 specific strategies --- affect, gender, identity, hedges, and negation. The first three are construct-driven strategies for sentiment, sexism, and hate speech, respectively, while the last two are construct-agnostic.\footnote{A construct-driven strategy for one construct could be construct-agnostic for another, e.g., changing affect words is a construct-agnostic strategy for sexism and hate speech.} 
We use a set of lexica for discerning each strategy --- a lexicon of positive and negative words for affect~\cite{hu2004mining}\footnote{obtained from: \url{https://www.cs.uic.edu/~liub/FBS/sentiment-analysis.html\#lexicon}}, list of gender words\footnote{obtained from: \url{https://github.com/uclanlp/gn_glove/tree/master/wordlist}} and a list of identity-based hateful terms and slurs~\cite{silva2016analyzing}\footnote{obtained from HateBase through their API}.
For negation, we use the list compiled by~\citet{ribeiro2020beyond} and for hedges, we use~\citet{islam2020lexicon}. Table~\ref{tab:cf_type_dist} enumerates the different types of CAD. We consider any counterfactual that does not fall under the construct-driven category to be construct-agnostic, e.g., 21\% of the CAD for sexism is construct-agnostic (as 79\% is construct-driven).

To determine whether a CAD sample is construct-driven or -agnostic, we first find the difference between the original datapoint and its counterfactually augmented counterpart and retrieve the additions and deletions based on that difference. We then check if the additions or deletions contain any of the words with the strategy-associated lexicon. Note that a single counterfactual example could span multiple strategies; e.g, the tweet ``It was \textbf{horrible}, I could \textit{not watch it}'', with the counterfactual ``It was \textbf{excellent}, I could \textit{watch it many times}'' pertains to a change in affect and negation. We sample 100 random original-counterfactual pairs over all constructs to validate our automatic categorization and find that for 89 cases, we are able to correctly label the annotation strategy. Errors include misplellings of slurs, or creative distancing strategies like ``[identity] stink'' to ``awful graffitti I saw today: `[identity] stink'{\thinspace}''.

\textbf{Models trained on different types of CAD. }We train models on the different types of counterfactuals (see Table~\ref{tab:cf_type_dist}). Specifically, we train three types of models: (a) models trained on just construct-driven counterfactuals (CF\_c); (b) models trained on just construct-agnostic counterfactuals (CF\_r); and (c) models trained on equal proportions of both (CF\_a).\footnote{We train the last type with equal proportions instead of a random set of CAD like the CF models in RQ1 and RQ3 since construct-driven CAD makes up the majority for sexism.} We measure the macro F1 of each of these types of models for the out-of-domain data. Since we have almost negligible construct-agnostic CAD for sentiment, we conduct the analyses for RQ2 on sexism and hate speech only.\footnote{One reason for the low proportion of construct-agnostic CAD in sentiment is the nature of the in-domain data; while for sexism and hate speech, the in-domain data consists of tweets or short single-sentence utterances, the data for sentiment comes from movie reviews which are much longer and have multiple edits made throughout. It is natural to find reviews which have a negation injected, while also having an affect word being changed. 
} Furthermore, due to less than 50\% CAD for certain types, instead of a 50\% injection, we vary the proportion between 10\% to 20\%.

\begin{figure}
    \centering
    \includegraphics[scale=0.25]{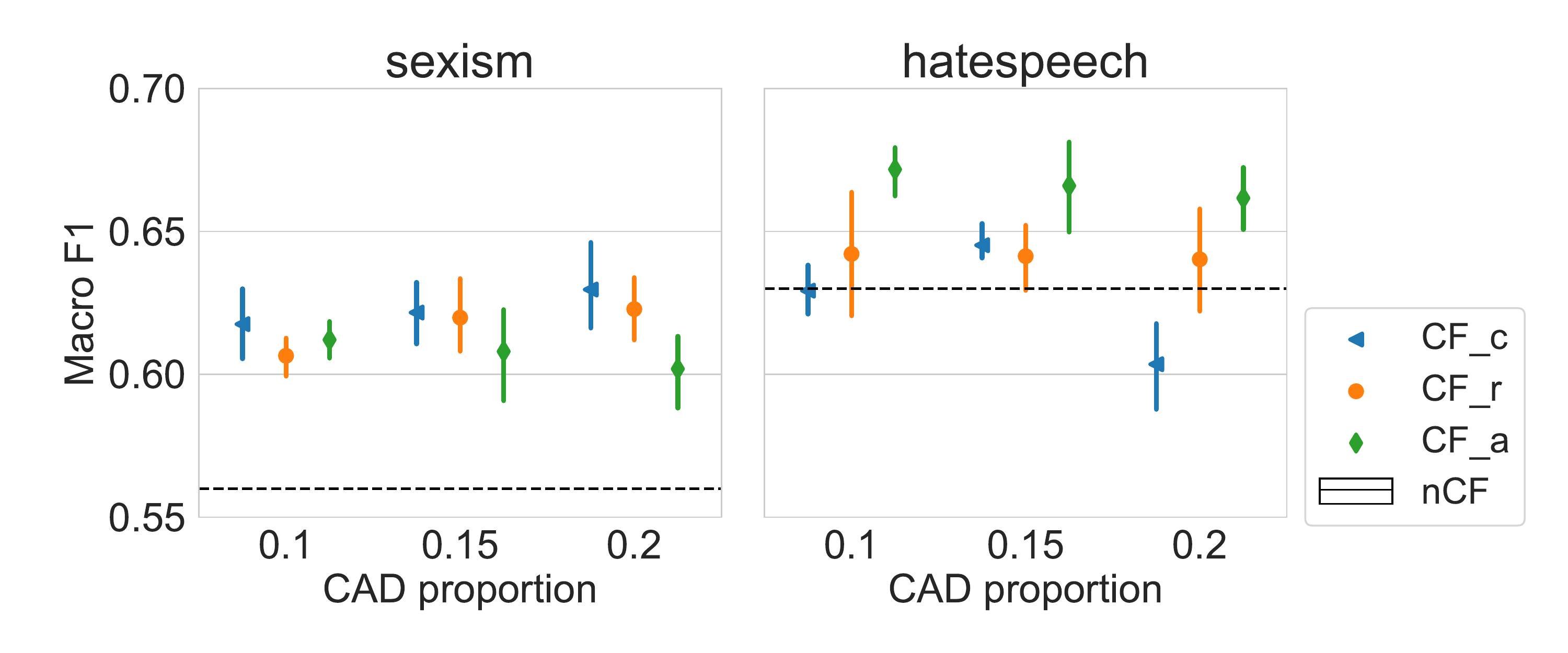}
    \caption{\textbf{Performance (macro F1) of BERT models trained on different types of CAD over different injection proportions on the out of domain data.} nCF model performance is included as a reference. Construct-driven CAD performs well especially for sexism, while in hate speech, diverse CAD is better.}
    \label{fig:counterfactuals_by_typology}
\end{figure}

\subsection{Results}

We show the macro F1 of these three types of models on out-of-domain data over different CAD proportions in Figure~\ref{fig:counterfactuals_by_typology}. We obtain mixed results for RQ2. First, we see that performance increases with the CAD proportion, except for hate speech at 20\% (complemented by our analysis in Appendix~\ref{app:injection}). \textit{Our results indicate that models trained on construct-driven CAD (CF\_c) are more effective than other types for sexism, especially at higher injection proportions}. On the other hand, for hate speech, CF\_a, or the diverse set of counterfactuals are better. Models trained on construct-agnostic CAD (CF\_r) have mixed efficacy.

\subsection{RQ3: Do models trained on CAD rely on fewer artifacts?}\label{sec:explanations}

\begin{figure*}[!t]
    \centering
    \includegraphics[scale=0.342]{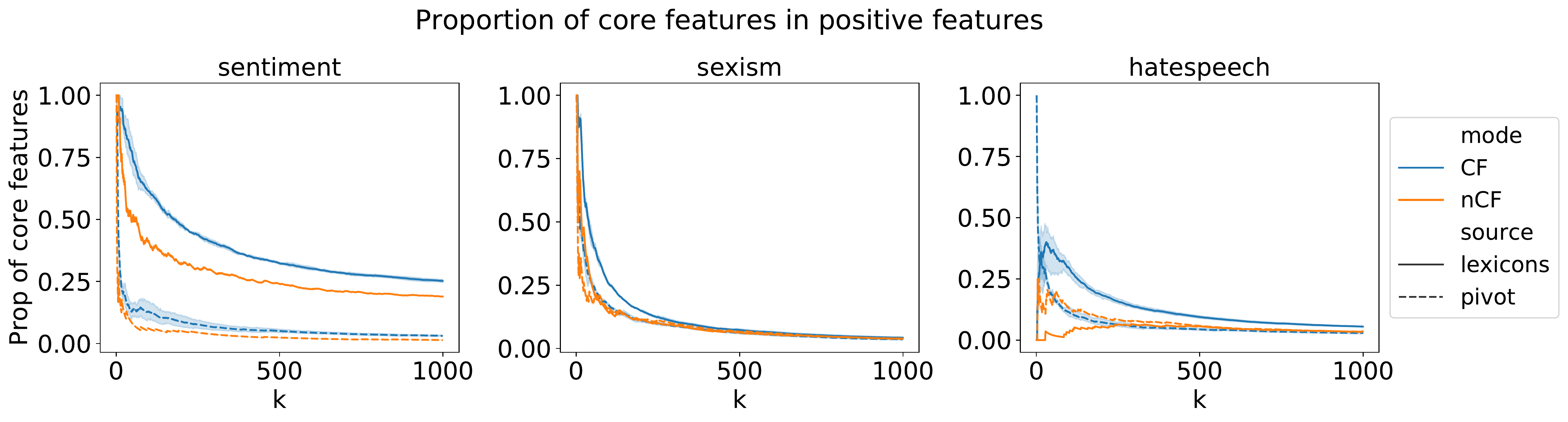}
    \caption{\textbf{Proportion of core features in the top-k positive LR global feature importances} Models trained on CAD have a higher proportion of core (non-spurious) features, demonstrated by the gap between CF and nCF models in lexica, especially for sentiment and hate speech. For pivot words, the gap is smaller.}
    \label{fig:global_features_lexica}
\end{figure*}

While the overall performance gains can help us understand the improvements led by counterfactual data, we still do not know how or why these performance gains came to fruition. To that end, we apply explainability techniques to shed light on the models' inner workings and pinpoint what changes were brought about by the counterfactual data.


While explainability for transformer models like BERT is an active area of research, explanation methods for them are usually at the level of individual predictions (local explanations). In this work, as we wish to assess how CAD holistically impacts social NLP models, we are primarily interested in model understanding over prediction understanding. Therefore, we need a way to aggregate local explanations into global features, a non-trivial task~\cite{van2019global}. Furthermore, explanations generated in an unsupervised way are not always faithful~\cite{atanasova2020diagnostic} and BERT does not learn weights for words, but for subwords,\footnote{see e.g. \url{https://huggingface.co/transformers/tokenizer_summary.html}} making it difficult to find the importance of words. Therefore, as we cannot ascertain the reliability of BERT-generated global features and since LR and BERT models show similar trends in overall performance, for this analysis, we use the built-in feature weights of the LR models to compute the top-k global important features for CF and nCF models. We experiment with BERT explanations and include the result in Appendix~\ref{app:bert_explain} but we leave a detailed analysis of aggregation strategies of local BERT explanations for future work.

\textbf{Quantitative Global Feature Analysis. }As the goal of training on CAD is to reduce the reliance on spurious features, we hypothesize that \textit{CF models have higher proportions of core (non-spurious) features in their feature ranking.} `Core' features are those that are consequential manifestations of the construct (e.g. the word `happy' for sentiment), while spurious features are those that happen to be correlated with the construct in a particular dataset while not being truly indicative of it (`movie' for sentiment). Therefore, core features of a particular construct span multiple domains or datasets of that construct. Besides manually inspecting the top-20 global features, we also quantitatively assess the presence of spurious features in the global feature importances, i.e., we check the proportion of core features in the top feature rankings. 

\textbf{Identifying core features. }To answer RQ3, we need a source of core features, or words associated with each of our constructs. To do so, we define two sources --- (a) \textit{lexica} and (b) \textit{pivot words}. For the first, we use the same lexica for understanding the construct-driven modification strategies in RQ2, i.e., affect words for sentiment, gender words for sexism, and identity-based hate words for hate speech. Note that, while for sentiment, we have a list of core features for both classes, for sexism and hate speech, we only have core features for the positive class for sexist and hate cases, and \textit{not} for non-sexist and non-hate cases. For the second source, we turn to the literature on domain adaptation, particularly work on \textit{pivot words}~\cite{blitzer2007biographies}. Concretely, for a given construct, we find words that are highly frequent in both domains; then find their correlation with the out-of-domain dataset labels to reduce the inclusion of in-domain artifacts. We rank these words based on mutual information and use the first 100 words as a set of core features. The list of pivot words is in Appendix \ref{app:pivot}.

\subsection{Results}

We manually inspect the top 20 features ranked most important by each model. The non-counterfactual models tend to learn more domain-specific features such as `script' (sentiment), `football' (sexism), and `wrong' (hate speech), which prevents them from generalizing to other domains. 
\textit{The counterfactual models show fewer spurious features in their most important features, instead having more affect words (sentiment), gendered words (sexism), and identity-based slurs (hate speech).} The top-20 features 
are in Appendix \ref{app:top_20}. 

To scale this analysis, we use lexica and pivot words as proxies for core, i.e, non-spurious features. We plot the proportion of core features in the top positive feature ranking. Figure~\ref{fig:global_features_lexica} shows that LR CF models rank core features more highly, especially based on the core feature list from lexica, strongly evident for sentiment, but also present to a lesser degree for sexism and hate speech. \textit{Therefore, our analysis indicates that training on CAD leads to reduced reliance on spurious features, while promoting core features.} In contrast to lexica words, for pivot words, the gap between CF and nCF models is much smaller for sentiment and sexism. Whereas, for hate speech, the nCF models tend to have a higher proportion of core pivot word features after a certain k. We include the results for proportion of negative features in Appendix \ref{app:neg_exp}.

\section{Related Work}

Our work connects the area of learning with counterfactuals to improve NLP models' robustness with the area of social NLP. 

\textbf{Counterfactuals in NLP. }
Counterfactuals in NLP have been used for model testing, and explanation, but in this work, we are interested in using them for training models. Counterfactuals can be used for augmenting training data where previous research, focused on sentiment and NLI, has shown models trained on this augmented data are more robust to data artifacts~\cite{kaushik2019learning,teney2020learning}. Counterfactuals need not always be label-flipping, but usually entail making minimal changes to original data either, and can be generated by manually or automatically~\cite{nie2019adversarial}. Recent work has also addressed automatic CAD generation through lexical or paraphrase changes~\cite{garg2019counterfactual,iyyer2018adversarial}, templates~\cite{nie2019adversarial}, and controlled text generation~\cite{wu2021polyjuice,madaan2020generate}. Concurrent and closely related to our work,~\citeauthor{joshi2021investigation} assess the efficacy of CAD for Natural Language Inference and Question Answering, and find that diverse CAD is crucial for improving generalizability, in line with our current work. On the other hand, CAD generated by human annotators has not been analyzed in detail to see which strategies are used for generating counterfactuals nor which strategies are more effective, particularly for social computing NLP tasks.

In this work, we focus on human generated, label-flipping counterfactuals for relatively understudied constructs in this domain --- sexism and hate speech, while more importantly focusing on how CAD impacts models. Inspired by causal mediation~\cite{pearl2014interpretation}, we put forth a typology of construct-driven and construct-agnostic CAD. Complementing previous research on overall performance, we take a deeper dive into which features CAD promotes, and which types are effective. 


\textbf{Social Computing and Online Abuse Detection.} Even though sentiment, sexism, and hate speech can all be considered social computing tasks, the latter two, and generally NLP tasks related to abuse detection~\cite{schmidt2017survey,jurgens2019just,nakov2021detecting,vidgen2020directions,sarwar2021flagging}, differ from tasks like sentiment and NLI because of their subjective nature and the relatively higher risk of social harms incurred by deploying spurious and non-robust models for decision making. Previous work has shed light on several dimensions of hate speech data that prevents generalisation, such as imprecise construct specification~\cite{samory2021call}, biased data collection~\cite{ousidhoum2020comparative}, and annotation artifacts~\cite{waseem-2016-racist}. Several solutions have been proposed for these issues such as adversarial data generation~\cite{dinan2019build}, dynamic benchmarking~\cite{kiela2021dynabench} and debiasing techniques~\cite{nozza2019unintended}. 

Building on these threads of research, we aim to understand the benefits of different types and proportions of CAD in training social NLP models.

\section{Discussion: Designing Counterfactually Augmented Data}

NLP models are now embedded in many real-world applications and understanding their limits and robustness is of the utmost importance, especially for social computing applications. In this work, using a detailed and systematic set of analyses we establish convergent validity of the use of counterfactually augmented data for improving the reliability of datasets, particularly for learning social constructs like sentiment, sexism, and hate speech. 

Through extensive testing on different types of data, including adversarial data, we corroborate and strengthen previous findings that training on CAD leads to robust models~\cite{kaushik2019learning,samory2021call}. While it is promising that CF models do not fall prey to adversarial perturbations any more than their nCF counterparts, the disparity in out-of-domain performance and the lack thereof in adversarial examples might indicate that adversarial examples are not strong testbeds for detecting model robustness on out-of-domain data. 

Having established 
this, we assessed if all CAD are equally effective. Using a fine-grained categorization of counterfactual generation strategy, we find that to not be the case, where for sexism, examples generated by directly acting on the construct are more effective in improving overall performance. Our results indicate that different strategies have different strengths, and model designers can prioritize certain strategies over others based on their needs. Finally, using explainability techniques, we establish that models trained on CAD tend to rely and promote core features over spurious ones using lexica and domain-agnostic words. 



\textbf{Limitations.} 
The main limitation of our paper is that we rely on lexica and automated methods for several prongs of our analyses --- for detecting core or non-spurious features and for classifying the different types of counterfactuals. Although manual vetting of both reveals that the results are sound, we caution against using them outside of this particular context. As we are limited in our computational resources, we further did not compare different explanation generation methods. 


The second limitation of our work is using explanations from a bag-of-words LR model, which is motivated by two factors. First, since we want to understand how counterfactuals affect ML models holistically, we require precise and faithful global explanations, making the feature importances from LR an ideal choice. Second, explanation methods are an active area of research for Transformer models, and aggregating local explanations to global ones remains challenging~\cite{van2019global}. As we could not guarantee that the aggregated BERT explanations would reflect the model's internal decision-making mechanism, we default to the LR models for this particular analysis.


\textbf{Future Work.} We used lexica to detect types of counterfactuals, however, they have several drawbacks such as limited recall. A supervised classification approach could be considered as a step forward, which might be more sophisticated and accurate. On the other hand, such an approach would have to grapple with the complexities of the task of finding types of counterfactuals, since the input is paired (original-counterfactual) rather than a single document. Furthermore, a labeled dataset of sufficient size and careful feature engineering would be needed, which could be tackled in future work.


The use of counterfactuals for training data augmentation is fairly recent, with work by~\citeauthor{kaushik2019learning} in 2019, even more so for social computing constructs. Therefore, there are several open questions about their properties as training data, including the notion of \textbf{minimality of a counterfactual}, i.e, what constitutes a minimal edit in generating CAD, either through quantitaive measures such as lexical distance, qualitative approaches, or their combination. Recent work has also attempted to automatically generate CAD~\cite{wu2021polyjuice,madaan2020generate}. However, comparing automated and human generated counterfactuals as training data is an open question and the analysis conducted in our work could be reused for this comparison.

Finally, the measurement of all three constructs in this work were modeled as binary classification tasks. Indeed, the counterfactual generation framework implicitly assumes binary labels as the approach asks for annotators to \textit{flip} the label. Nevertheless, social constructs are multifaceted and could be modeled as multiclass (or even, multilabel) classification tasks. Future work could extend the current binary setup of counterfactual generation to accommodate multiclass classifications for example, through a one-vs-rest approach.

\section{Conclusion}

We take a deeper dive into the utility of training on counterfactually augmented data (CAD) for improving the robustness of social NLP models. For three text classification constructs---sentiment, sexism, and hate speech---we train LR and BERT models with and without counterfactual data. For the counterfactual models, we experiment with different sampling strategies to understand how different types of CAD affect model performance. 
Firstly, we corroborate previous findings on using counterfactual data, showing that models trained on CAD have higher out-of-domain performance. 
Our work's core novelty is that we study different strategies for CAD generation, and find that examples generated by acting on the construct are effective for sexism, while a diverse set is better for hate speech. Finally, 
we show that models trained on CAD promote core or non-spurious features over spurious ones. Taken together, our analysis serves as a blueprint for assessing the potential of CAD, while our findings can help dataset and model designers design better CAD for social NLP tasks.

\section*{Acknowledgments}

We thank members of the CopeNLU group and the anonymous reviewers for their constructive feedback. Isabelle Augenstein's research is partially funded by a DFF Sapere Aude research leader grant.

\input{ethics}

\bibliography{mypaper}
\bibliographystyle{acl_natbib}

\newpage
\input{appendix_arxiv}

\end{document}

%% file: ethics.tex
\section{Ethical Considerations}

In this work, we attempt to understand the connection between training on counterfactually augmented data and increased model robustness. Our work centers on social NLP constructs like sexism and hate speech, whose manifestations in data can be harmful and potentially traumatizing to researchers. Furthermore, the sensitive nature of this data has the potential of victimising or re-victimizing the people referred to in them. Therefore, in accordance with ethical guidelines~\cite{vitak2016beyond,zimmer2017internet,vidgen2020directions} we conduct our analyses on aggregate data only and do not infer any attributes of the speakers in the data. We release a dataset which only contains the IDs of the original data and the typology labels we annotate.

Following common practice in NLP, we use a gendered lexicon that only contains gendered words based on the gender binary. We acknowledge that this practice is exclusionary towards non-binary individuals. We alleviate this to a certain extent by having a broader and more detailed list of identity terms, which also contains hateful terms and slurs directed towards non-binary people. In future, we hope to adopt a more intersectional perspective which is more inclusive of the sexism faced by trans and non-binary people~\cite{serano2016whipping,winter2009transpeople}. 

Constructs like sexism and hate speech detection are often depicted as neutral or objective but they are deeply contextual, subjective and ambiguous~\cite{vidgen2019challenges,jurgens2019just,nakov2021detecting}, where misclassifications can cause harm~\cite{blackwell2017classification}. We use lexica to determine core features of sexism or hate speech, but we acknowledge that both of these may manifest in context-dependent ways and there is no single objective determinant of hate speech or sexism (or even sentiment). Furthermore, promoting features like identity terms can increase the risk of misclassifying non-hate content with such terms, such as disclosure or reports of facing hate speech, leading to unintended bias~\cite{blodgett2020language}. 

We do not undertake any further data generation or data annotation by human subjects, as we use data made available by previous researchers and use lexica for annotating counterfactual types. Nonetheless, as we show the potential of CAD in improving some aspects of model robustness, we hope that the community will adopt annotation guidelines that factor in the risk of harm that annotators and CAD designers working on abusive language might face~\cite{vidgen2020directions}. 

We aim to understand how CAD improves model robustness, but we acknowledge and caution that these types of data augmentation can also be used to poison NLP models and cause them to have several harmful properties~\cite{wallace2020customizing,sun2018data}.

%% file: appendix_arxiv.tex
\section*{Appendix}

Here is the appendix for our paper, ``How Does Counterfactually Augmented Data Impact Models for Social Computing Constructs?''.  The appendix contains details for facilitating reproducibility (\ref{app:rep}), the LR results to supplement the BERT results in the paper (\ref{app:lr_results}), the entire list of pivot words (\ref{app:pivot}), global top-20 features (\ref{app:top_20}), results for negative features' in RQ3 (\ref{app:neg_exp}), and the BERT explanations (\ref{app:bert_explain}).

\textbf{Caution: The appendix contains examples of terminology found to be discerning of hate speech and sexism, and are therefore, of an offensive nature.}

\section{Reproducibility}\label{app:rep}

\subsection{Compute Infrastructure}

All models were trained or finetuned on a 40 core Intel(R) Xeon(R) CPU E5-2690 (without GPU).

\subsection{Model Training Details: Hyperparameters and Time Taken}\label{app:hyper}

We preprocess all the data by removing social media features such as hashtags and mentions. The hyperparameter bounds for LR models are: 

\begin{enumerate}
    \item stopwords: English, none, English without negation words 
    \item norm: ('l1', 'l2')
    \item C: (0.01, 0.1, 1)
    \item penalty: ('l2', 'l1')

\end{enumerate}

while for BERT we use: 

\begin{enumerate}

    \item epochs:[4, 5]
    \item learning rate: 2e-5, 3e-5, 5e-5

\end{enumerate}

\begin{table*}[]
\small
\centering
\begin{tabular}{@{}llll@{}}
\toprule
construct                    & model    & \begin{tabular}[c]{@{}l@{}}best model \\ hyperparameters\end{tabular} & \begin{tabular}[c]{@{}l@{}}time to train \\ (one run)\end{tabular} \\ \midrule
\multirow{4}{*}{sentiment}   & CF LR    & english without negation, l1, 1, l1                                   & 24.12s                                                             \\
                             & CF BERT  & epochs: 5, learning rate: 5e-5                                        & 4h07m32s                                                           \\
                             & nCF LR   & english without negation, l1, 0.1, l1                                 & 26.88s                                                             \\
                             & nCF BERT & epochs: 5, learning rate: 3e-5                                        & 4h10m20s                                                           \\
\midrule                             
\multirow{4}{*}{sexism}      & CF LR    & english, l2, 0.01, l2                                & 5.42s                                                              \\
                             & CF BERT  & epochs: 5, learning rate: 2e-5                                        & 3h42m20s                                                           \\
                             & nCF LR   & none, l2, 0.01, l2                                & 4.87s                                                              \\
                             & nCF BERT & epochs: 5, learning rate: 2e-5                                        & 3h38m57s                                                           \\
\midrule                             
\multirow{4}{*}{hate speech} & CF LR    & english without negation, l2, 0.01, l2                                & 26.27s                                                             \\
                             & CF BERT  & epochs: 4, learning rate: 5e-5                                        & 17h54m03s                                                          \\
                             & nCF LR   & english without negation, l2, 0.01, l2                                & 26.67s                                                             \\
                             & nCF BERT & epochs: 5, learning rate: 5e-5                                        & 17h39m29s                                                          \\ \bottomrule
\end{tabular}
\caption{Hyperparameters for CF (trained on 50\% CAD) and nCF models.}
\label{tab:hypers}
\end{table*}

\begin{table*}[]
\centering
\small
\begin{tabular}{@{}llll@{}}
\toprule
construct                    & model      & best model hyperparams                & time to train (one run) \\ \midrule
\multirow{6}{*}{sentiment}   & CF\_c LR   & english, l1, 1, l1                    & 25.53s                  \\
                             & CF\_a LR   & none, l1, 0.1, l1                     & 23.69s                  \\
                             & CF\_r LR   & none, l1, 0.1, l1                     & 26.88s                  \\
                             & CF\_c BERT & epochs: 5, learning rate: 3e-5        & 4h10m20s                \\
                             & CF\_a BERT & epochs: 5, learning rate: 3e-5        & 4h21m05s                \\
                             & CF\_r BERT & epochs: 5, learning rate: 3e-5        & 4h11m02s                \\
                             \midrule
\multirow{6}{*}{sexism}      & CF\_c LR   & english, l1, 1, l1                    & 5.91s                   \\
                             & CF\_a LR   & english without negation, l1, 1, l1   & 6.15s                   \\
                             & CF\_r LR   & english, l2, 0.1, l2                  & 5.27s                   \\
                             & CF\_c BERT & epochs: 5, learning rate: 5e-5        & 3h42m20s                \\
                             & CF\_a BERT & epochs: 5, learning rate: 3e-5        & 3h34m36s                \\
                             & CF\_r BERT & epochs: 5, learning rate: 2e-5        & 3h50m18s                \\
                             \midrule
\multirow{6}{*}{hate speech} & CF\_c LR   & english without negation, l1, 1, l1   & 33.35s                  \\
                             & CF\_a LR   & english without negation, l1, 0.1, l1 & 30.08s                  \\
                             & CF\_r LR   & none, l1, 0.1, l1                     & 32.67s                  \\
                             & CF\_c BERT & epochs: 5, learning rate: 3e-5        & 18h09m11s               \\
                             & CF\_a BERT & epochs: 5, learning rate: 3e-5        & 17h58m33s               \\
                             & CF\_r BERT & epochs: 5, learning rate: 2e-5        & 17h49m46s               \\ \bottomrule
\end{tabular}
\caption{CF models trained on different types of CAD.}
\label{hypers_CAD}
\end{table*}

For LR, we have 36 combinations over 5 fold cross-validation, leading to 180 fits, while for BERT, we have 6 combinations also over 5 fold CV, leading to 30 fits. 

We use gridsearch for determining hyperparameter, where the metric for selection was macro F1.  
Run times and hyperparameter configuartions for the best performance for all CF (with randomly sampled 50\% data) and nCF models (RQ1) are included in Table~\ref{tab:hypers}. The hyperparameters and run times for the CF models trained on different types of CAD (RQ2) are in Table~\ref{hypers_CAD}.

\subsection{Metrics} 

The evaluation metrics used in this paper are macro average F1, positive class precision for RQ1 and RQ2. We used the sklearn implementation of these metrics: \url{https://scikit-learn.org/stable/modules/generated/sklearn.metrics.precision_recall_fscore_support.html}.  For RQ3, we compute the fraction of core features in a feature list based on intersection with the lexica and the pivot words (included in the appendix~\ref{app:pivot}). The code for computing the metric is included in our code (uploaded with the submission)

\subsection{Model Parameters}

Model parameters are included in Table~\ref{tab:model_parameters}.

\begin{table}[]
\begin{tabular}{@{}lll@{}}
\toprule
construct   & model    & \#params              \\ \midrule
Sentiment   & CF LR    & 16282                 \\
            & nCF LR   & 18478                 \\
            & CF BERT  & \multirow{2}{*}{110M} \\
            & nCF BERT &                       \\
\midrule            
Sexism      & CF LR    & 4750                  \\
            & nCF LR   & 5505                  \\
            & CF BERT  & \multirow{2}{*}{110M} \\
            & nCF BERT &                       \\
\midrule            
Hate speech & CF LR    & 13763                 \\
            & nCF LR   & 14800                 \\
            & CF BERT  & \multirow{2}{*}{110M} \\
            & nCF BERT &                       \\ \bottomrule
\end{tabular}
\caption{Number of model parameters for the CF and nCF models.}
\label{tab:model_parameters}
\end{table}

\section{LR Results}\label{app:lr_results}

Here we show the results for LR models. While the BERT models have much higher performance than LR, both family of models show similar trends, indicating that CAD is beneficial across model families. We show the results for LR for adversarial examples in Table~\ref{tab:performance_adv_lr}. We also experiment with different proportions of CAD and measure their effect on performance in Figure~\ref{fig:injection_proportion}. Finally, we also include the performance of the LR models trained on different types of CAD in Figure~\ref{fig:counterfactuals_by_typology_lr}. 

\subsection{Injection Analysis. }\label{app:injection}

\begin{figure}
    \centering
    \includegraphics[scale=0.184]{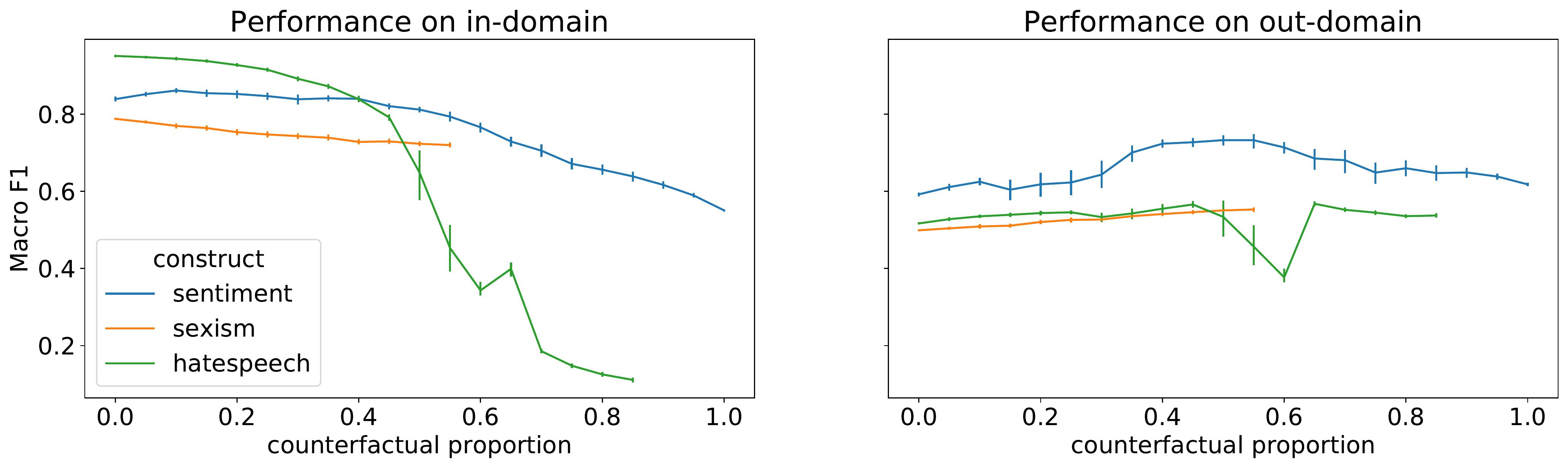}
    \caption{\textbf{Performance of LR models trained on different proportions of counterfactually augmented data over 5 runs.} For all three constructs, we see that models degrade consistently in in-domain datasets, while improve to a certain point for out-of-domain data.}
    \label{fig:injection_proportion}
\end{figure}

\begin{figure}[!htbp!]
    \centering
    \includegraphics[scale=0.25]{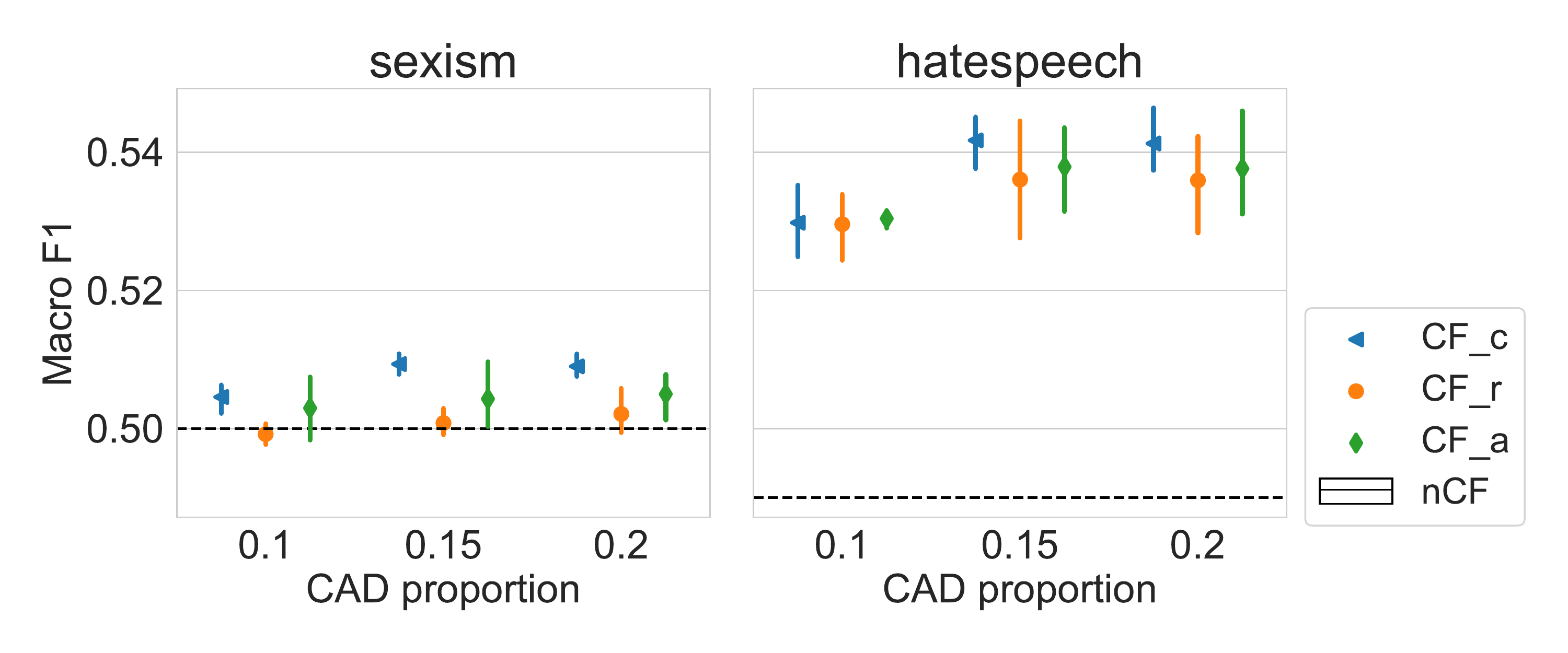}
    \caption{\textbf{Performance (macro F1) of LR models trained on different types of counterfactually augmented data over different injection proportions on the out of domain data.} Construct-driven CAD performs well especially for sexism (like the BERT models), while in hate speech there is more variance.}
    \label{fig:counterfactuals_by_typology_lr}
\end{figure}

In the main paper, we have replaced half of the original data with CAD (25\% for sexism) and seen that it improves out-of-domain performance. But is there a limit to CAD's benefits? We investigate which amount of counterfactually augmented data is effective. We assess how different proportions of counterfactual examples injected affect the overall performance in Figure~\ref{fig:injection_proportion}. While substituting original training data with counterfactually augmented data leads to reduced performance in-domain where the decrease is proportional to the amount of counterfactually augmented data, the trends are dissimilar for out-of-domain performance. Models trained on counterfactually augmented data perform better out-of-domain \textbf{but only} to a certain extent, after which point they begin degrading, potentially due to learning CAD-specific cues, though the limits are different for different constructs. \textbf{Our analysis implies that while injecting counterfactually augmented data can be indeed effective for out-of-domain data, using an equal proportion of counterfactual and normal data achieves best performance.}

\begin{table}[]
\small
\begin{tabular}{@{}lllll@{}}
\toprule
            &        &                    & \multicolumn{2}{l}{Macro F1} \\ \midrule
mode        &        &                    & CF            & nCF          \\
\midrule
construct   & method & dataset            &               &              \\
\midrule
sentiment   & logreg & adv\_inv           & 0.80          & 0.85         \\
sentiment   & logreg & adv\_inv original  & 0.82          & 0.86         \\
sentiment   & logreg & adv\_swap          & 0.75          & 0.83         \\
sentiment   & logreg & adv\_swap original & 0.82          & 0.86         \\
\midrule
sexism      & logreg & adv\_inv           & 0.71          & 0.76         \\
sexism      & logreg & adv\_inv original  & 0.71          & 0.77         \\
sexism      & logreg & adv\_swap          & 0.68          & 0.75         \\
sexism      & logreg & adv\_swap original & 0.72          & 0.78         \\
\midrule
hate speech & logreg & adv\_inv           & 0.75          & 0.92         \\
hate speech & logreg & adv\_inv original  & 0.75          & 0.91         \\
hate speech & logreg & adv\_swap          & 0.66          & 0.86         \\
hate speech & logreg & adv\_swap original & 0.73          & 0.92         \\ \bottomrule
\end{tabular}
\caption{\textbf{The Performance of LR models on adversarial data.}}
\label{tab:performance_adv_lr}
\end{table}

\section{Pivot Words}\label{app:pivot}

Here are the list of pivot words per construct. Not all pivot words are meaningfully representative of the construct and contain out-of-domain artifacts like `elvis' and `south'. Since none of the models were trained on out-of-domain data, we do not expect such words to inflate our metrics in Figure 2 of the main paper.

\textbf{Sentiment.} 'long', 'boring', 'never', 'glad', 'see', 'ending', 'credits', 'roll', 'not', 'good', 'buy', 'watch', 'someone', 'head', 'like', 'elvis', 'real', 'king', 'movie', 'bad', 'time', 'worst', '7', 'throughout', 'something', 'anything', 'really', 'waste', 'garbage', 'spanish', 'smart', 'interesting', 'stories', 'case', 'name', 'badly', 'missed', 'chance', 'story', 'seen', 'movies', '39', 'major', 'release', 'span', 'awful', 'unhappy', 'complete', 'b', 'instead', 'classic', 'terrible', 'acting', 'film', 'watched', 'unless', 'looking', 'cure', 'insomnia', 'imagine', 'anyone', 'actually', 'thinking', 'best', 'given', 'ever', 'top', 'direction', 'great', 'got', 'turned', 'silly', 'shame', 'idea', 'potential', 'shot', 'lots', 'example', 'dr', 'daughter', 'ages', 'years', 'wait', 'video', 'much', '100', 'brain', 'cell', 'killing', 'way', 'money', 'store', 'mad', 'sat', 'spent', 'absolutely', 'slow', 'wish', 'could', 'say'

\textbf{Sexism.} 'fuck', 'women', 'shit', 'web', 'experience', 'similar', 'key', 'know', 'twitter', 'additional', 'controls', 'verified', 'man', 'hungry', 'making', 'best', 'damn', 'sandwich', 'ever', 'limit', 'michelle', 'obama', 'happy', 'looked', 'beautiful', 'deep', 'blue', 'purple', 'dress', 'wore', 'today', 'knock', 'found', 'color', 'made', 'black', 'street', 'player', 'monkey', 'started', 'getting', 'heat', 'obviously', 'logical', 'state', 'family', 'paid', 'father', 'mother', 'lost', 'stand', 'watching', 'conceited', 'idiots', 'husband', 'right', 'expect', 'wife', 'times', 'think', 'less', 'clearly', 'men', 'emotional', 'believe', 'wear', 'dresses', 'trying', 'decide', 'time', 'contact', 'police', 'call', 'w', 'lawyer', 'never', 'thought', 'say', 'unless', 'trouble', 'involved', 'actually', 'girl', 'not', 'sensitive', 'seen', 'cabinet', 'gender', 'matters', 'dear', 'sexist', 'get', 'like', 'nagging', 'work', 'society', 'culture', 'giving', 'due', 'respect'

\textbf{Hatespeech.} 'burden', 'society', 'many', 'b', 'l', 'c', 'k', 'country', 'not', 'around', 'like', 'hate', 'called', 'nigger', 'horrible', 'people', 'smell', 'dirty', 'stinky', 'lazy', 'black', 'trans', 'fucking', 'hell', 'life', 'real', 'cunt', 'absolutely', 'muslims', 'street', 'cute', 'gay', 'non', 'immigrants', 'men', 'asian', 'women', 'living', 'area', 'really', 'nice', 'get', 'tired', 'time', 'foreigners', 'never', 'wash', 'sorry', 'bad', 'way', 'clever', 'blacks', 'kept', 'aside', 'actually', 'possible', 'seem', 'less', 'belong', 'south', 'east', 'refugees', 'general', 'would', 'rather', 'near', 'better', 'statistics', 'show', 'number', 'lack', 'work', 'hard', 'bring', 'world', 'made', 'lot', 'muslim', 'friends', 'since', 'mentally', 'retarded', 'contribute', 'anything', 'normal', 'banned', 'schools', 'apart', 'kids', 'soooo', 'much', 'killed', 'go', 'ahead', 'nuts', 'one', 'day', 'stop', 'getting', 'told'

\section{Top 20 words by construct for CF and nCF models}\label{app:top_20}

\begin{table*}[!htbp]
\centering
\begin{tabular}{@{}llrlrlrlr@{}}
\toprule
   & \multicolumn{4}{c}{Counterfactual}               & \multicolumn{4}{c}{Non-Counterfactual}           \\ \midrule
   & pos feature & pos coef & neg feature & neg coeff & pos feature & pos coef & neg feature & neg coeff \\ \midrule
0  & hilarious   & 0.77     & bad         & -3.45     & \color{red}{lives}       & 0.94     & bad         & -3.63     \\
1  & \color{red}{every}       & 0.81     & worst       & -3.23     & classic     & 0.99     & \color{red}{horror}      & -2.48     \\
2  & nice        & 0.82     & terrible    & -3.10     & amazing     & 1.00     & worst       & -2.33     \\
3  & loved       & 0.84     & boring      & -3.09     & \color{red}{young}       & 1.01     & boring      & -1.96     \\
4  & beautiful   & 0.88     & not         & -2.85     & \color{red}{romance}     & 1.02     & waste       & -1.95     \\
5  & funny       & 0.96     & awful       & -2.39     & \color{red}{highly}      & 1.04     & awful       & -1.87     \\
6  & interesting & 1.03     & poor        & -1.86     & loved       & 1.04     & terrible    & -1.85     \\
7  & brilliant   & 1.13     & poorly      & -1.65     & \color{red}{family}      & 1.09     & poor        & -1.62     \\
8  & awesome     & 1.18     & dull        & -1.63     & beautiful   & 1.11     & worse       & -1.54     \\
9  & perfect     & 1.38     & worse       & -1.54     & enjoyed     & 1.17     & plot        & -1.35     \\
10 & fantastic   & 1.39     & waste       & -1.53     & \color{red}{especially}  & 1.19     & stupid      & -1.35     \\
11 & exciting    & 1.44     & stupid      & -1.48     & fun         & 1.37     & horrible    & -1.32     \\
12 & \color{red}{well}        & 1.54     & horrible    & -1.46     & \color{red}{life}        & 1.42     & \color{red}{script}      & -1.27     \\
13 & love        & 1.66     & lame        & -1.37     & perfect     & 1.47     & \color{red}{like}        & -1.18     \\
14 & good        & 1.96     & weak        & -1.24     & best        & 1.48     & poorly      & -1.17     \\
15 & excellent   & 1.98     & nothing     & -1.17     & excellent   & 1.49     & money       & -1.15     \\
16 & wonderful   & 1.98     & fails       & -1.17     & wonderful   & 1.66     & don         & -1.14     \\
17 & amazing     & 2.20     & hate        & -1.11     & romantic    & 1.86     & pointless   & -1.12     \\
18 & best        & 2.33     & avoid       & -1.11     & love        & 1.97     & \color{red}{minutes}     & -1.11     \\
19 & great       & 4.38     & mediocre    & -1.06     & great       & 3.28     & \color{red}{movie}       & -1.07     \\ \bottomrule
\end{tabular}
\caption{We enumerate the top 20 global feature importances for sentiment detection. Spurious features are marked in red. We find that the counterfactual models learn more general less spurious or in-domain-specific features such as movie review related words.}
\label{tab:feature_importances}
\end{table*}

\begin{table*}
\centering
\begin{tabular}{@{}llrlrlrlr@{}}
\toprule
   & \multicolumn{4}{c}{Counterfactual}               & \multicolumn{4}{c}{Non-Counterfactual}           \\ \midrule
{} & pos feature &  pos coef & neg feature &  neg coeff & pos feature &  pos coef & neg feature &  neg coeff \\
\midrule
0  &       wifes &      1.24 &      people &      -3.37 &          \color{red}{in} &      1.48 &        love &      -2.28 \\
1  &    \color{red}{football} &      1.25 &        love &      -2.34 &        lady &      1.50 &      people &      -1.97 \\
2  &         boy &      1.33 &      person &      -1.99 &          \color{red}{me} &      1.62 &          as &      -1.78 \\
3  &      family &      1.34 &      adults &      -1.74 &     shouldn &      1.68 &       these &      -1.63 \\
4  &         \color{red}{not} &      1.34 &       adult &      -1.68 &        \color{red}{than} &      1.75 &        this &      -1.62 \\
5  &         sex &      1.55 &        kids &      -1.67 &         don &      1.79 &        that &      -1.61 \\
6  &        guys &      1.85 &      rookie &      -1.57 &      sexist &      1.91 &         kat &      -1.50 \\
7  &        wife &      1.90 &    grownups &      -1.54 &        \color{red}{when} &      2.10 &        same &      -1.50 \\
8  &     husband &      2.00 &      racist &      -1.45 &        girl &      2.14 &     without &      -1.38 \\
9  &        male &      2.05 &       happy &      -1.40 &      \color{red}{sports} &      2.17 &         you &      -1.30 \\
10 &        lady &      2.06 &     grownup &      -1.38 &     females &      2.29 &       lucky &      -1.22 \\
11 &     females &      3.12 &      elders &      -1.36 &       woman &      2.39 &       those &      -1.18 \\
12 &       girls &      3.59 &         kid &      -1.36 &         man &      2.48 &       happy &      -1.18 \\
13 &      sexist &      3.68 &       lucky &      -1.33 &       girls &      2.59 &        hope &      -1.17 \\
14 &         man &      3.92 &     without &      -1.33 &      should &      2.69 &          we &      -1.14 \\
15 &        girl &      3.92 &     freedom &      -1.25 &          \color{red}{rt} &      2.72 &       andre &      -1.08 \\
16 &      female &      4.29 &       equal &      -1.22 &    \color{red}{football} &      2.98 &        well &      -1.06 \\
17 &       woman &      4.31 &     changed &      -1.21 &      female &      3.28 &     equally &      -1.05 \\
18 &         men &      4.60 &       elder &      -1.16 &       women &      3.67 &        free &      -1.01 \\
19 &       women &      6.08 &        hope &      -1.10 &         men &      3.86 &       equal &      -0.95 \\
\bottomrule
\end{tabular}
\caption{We enumerate the top 20 global feature importances for sexism detection. Spurious features are marked in red. We find that the counterfactual models learn more general less spurious or in-domain-specific features such as movie review related words. Note that we only mark the spurious positive features because it is difficult to ascertain spurious features for the negative class. However, we see domain-specific terms like `Kat' and `Andre' in the top features as well.}
\label{tab:feature_importances_sexism}
\end{table*}

\begin{table*}[!htbp]
\centering
\begin{tabular}{@{}llrlrlrlr@{}}
\toprule
   & \multicolumn{4}{c}{Counterfactual}               & \multicolumn{4}{c}{Non-Counterfactual}           \\ \midrule
{} & pos feature &  pos coef & neg feature &  neg coeff & pos feature &  pos coef & neg feature &  neg coeff \\
\midrule
0  &       black &      2.29 &       racists &      -2.74 &        jews &      2.69 &     fucking &      -9.87 \\
1  &        wogs &      2.30 &          food &      -1.98 &        \color{red}{nice} &      2.76 &         lot &      -5.41 \\
2  &        fags &      2.32 &          cars &      -1.88 &       \color{red}{wrong} &      2.81 &        fuck &      -5.05 \\
3  &    trannies &      2.35 &        corona &      -1.84 &        urgh &      2.81 &   neighbour &      -4.63 \\
4  &     mussies &      2.42 &        racist &      -1.77 &    \color{red}{concepts} &      2.95 &        know &      -4.26 \\
5  &  foreigners &      2.47 &         awful &      -1.68 &     already &      2.98 &       sucks &      -4.23 \\
6  &       women &      2.48 &  supremacists &      -1.67 &       think &      2.99 &        call &      -4.18 \\
7  &      chinks &      2.49 &        strong &      -1.67 &        idea &      3.03 &       bitch &      -4.15 \\
8  &         jew &      2.49 &         covid &      -1.57 &       happy &      3.09 &        many &      -3.70 \\
9  &      whores &      2.52 &          cats &      -1.52 &         let &      3.53 &      friend &      -3.64 \\
10 &        jews &      2.53 &        hatred &      -1.51 &  \color{red}{understand} &      3.58 &      called &      -3.49 \\
11 &       white &      2.53 &          2020 &      -1.45 &       \color{red}{likes} &      3.60 &        area &      -3.45 \\
12 &      jewish &      2.59 &         fight &      -1.43 &         \color{red}{one} &      3.61 &        like &      -3.21 \\
13 &  immigrants &      2.64 &    homophobes &      -1.43 &        \color{red}{tell} &      3.76 &     useless &      -2.95 \\
14 &       camel &      2.66 &          dogs &      -1.43 &        \color{red}{talk} &      4.45 &        hate &      -2.90 \\
15 &       pakis &      2.72 &        tories &      -1.40 &       \color{red}{loves} &      4.76 &       black &      -2.76 \\
16 &        yids &      2.73 &          hear &      -1.40 &       women &      4.82 &      corona &      -2.71 \\
17 &        paki &      2.95 &          ukba &      -1.39 &    \color{red}{everyone} &      5.38 &       piece &      -2.69 \\
18 &     niggers &      3.00 &        haters &      -1.38 &        rude &      7.81 &         man &      -2.62 \\
19 &      blacks &      4.10 &         foxes &      -1.35 &        \color{red}{love} &     11.32 &     failure &      -2.57 \\
\bottomrule
\end{tabular}
\caption{We enumerate the top 20 global feature importances for hate speech detection. Spurious features are marked in red. We find that the counterfactual models learn less spurious or in-domain-specific features. Note that we only mark the spurious positive features because it is difficult to ascertain spurious features for the negative class.}
\label{tab:feature_importances_hate speech}
\end{table*}

\section{LR Negative Class Features}\label{app:neg_exp}

\begin{figure*}
    \centering
    \includegraphics[scale=0.344]{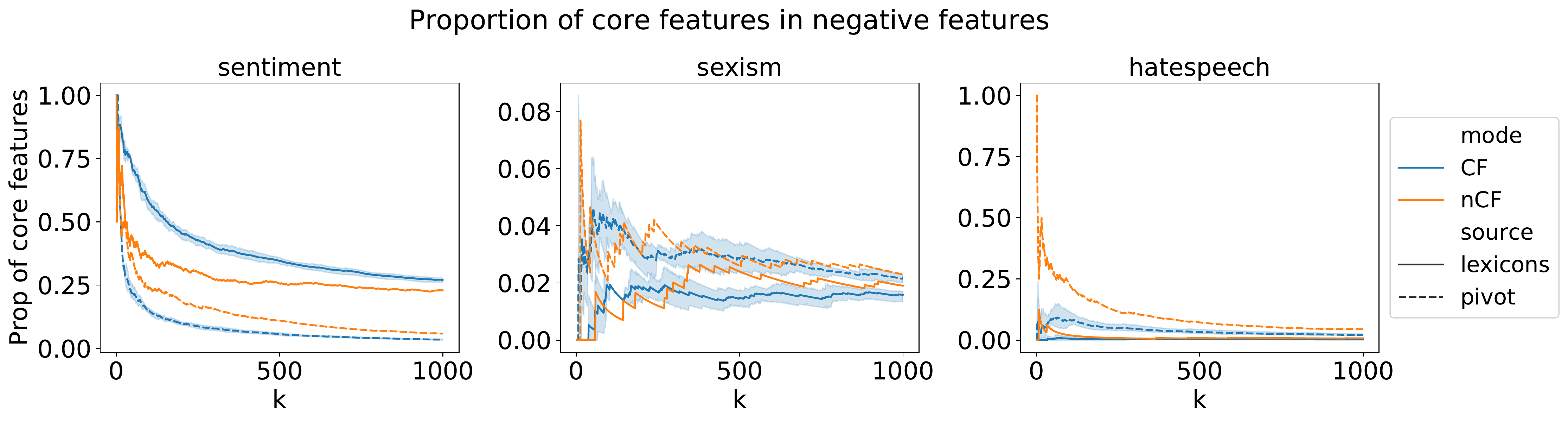}
    \caption{\textbf{Proportion of core features in the top-k negative global feature importances obtained by aggregating the local explanations for LR based on lexica and pivot words.} Unlike the case of positive features, sexism and hate speech do not show any clear trends in the proportion of core features. This contrasts them with sentiment, where there is an unclear of notion of salient features of not-hate and non-sexist.}
    \label{fig:global_features_lexica_neg}
\end{figure*}

Complementing Figure 2 in the main paper, we plot the proportion of core features in the most important \textit{negative} feature importance ranking of the LF CF ad nCF models in Figure~\ref{fig:global_features_lexica_neg}. This analysis demonstrates an interesting distinction between sentiment and the other two constructs, also seen in the top-20 global feature importances. Since it is difficult to envision negative features for constructs like sexism and hate speech, there is very little difference in the rankings of global features for the negative class between the two types of models, as opposed to sentiment where there is a clear difference between CF and nCF models. 

\section{BERT Explanations}\label{app:bert_explain}

As we state in the main paper, we use LR feature weights for understanding if CF models tend to rely on less spurious features. The reason for using LR is the purported unreliability of Transformer-based methods' explanations~\cite{jain2019attention,atanasova2020diagnostic} and the issues in aggregating local BERT explanations to global explanations for model understanding~\cite{van2019global}. Since not all explainability methods, especially for deep learning, are faithful. 

As an exploratory step, we complement the LR explanations with explanations for BERT, using \textit{Integrated Gradients}~\cite{sundararajan2017axiomatic}, where input importance is measured using the gradients computed with respect to the inputs. Previous research has found gradient-based methods outperform perturbation or model simplification-based approaches. As we are interested in model understanding rather than prediction understanding, we convert local explanations for BERT into a global feature ranking by aggregating the weights for every token in a local explanation.\footnote{Unlike LR where the weight for a token is fixed for all examples, BERT weighs tokens based on context.} While the trends are similar for sexism and sentiment, though the disparity between CF and nCF models is much smaller compared to the LR results, we caution against making concrete inferences from these results due to the potential unreliability of global BERT explanations.

\begin{figure*}
    \centering
    \includegraphics[scale=0.344]{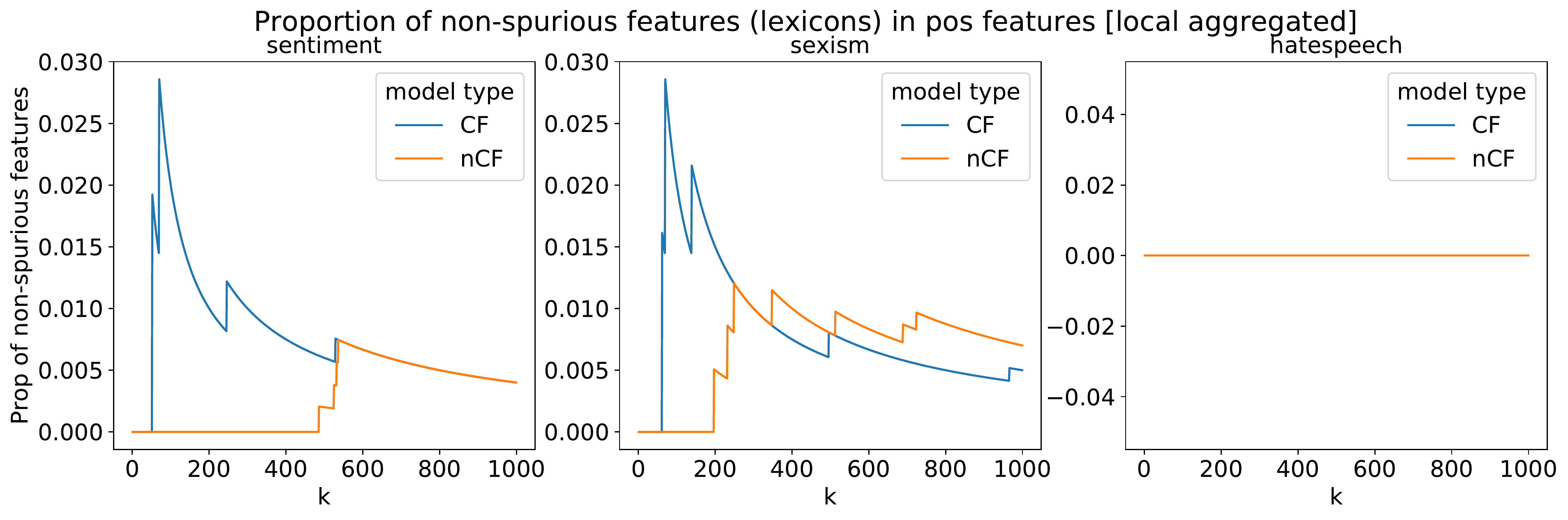}
    \caption{\textbf{Proportion of core features in BERT explanations}. BERT explanations are generated using gradients. While the results for sentiment and sexism are similar to LR explanation results.}
    \label{fig:bert_explain}
\end{figure*}